\documentclass[12pt]{article}
\usepackage{latexsym}
\usepackage{amsmath}
\usepackage{amssymb}
\usepackage{amsthm}
\usepackage{dsfont}
\usepackage{longtable}
\usepackage{color}
\usepackage{verbatim}
\usepackage[sort&compress,numbers]{natbib}
\usepackage{graphicx}
\usepackage{epstopdf}
\usepackage{float}
\usepackage{wrapfig}

\begin{document}

\title{New version of pseudo-Hermiticity in the two-sided deformation of Heisenberg
algebra}

\author{A.M. Gavrilik$^\sharp$ and I.I. Kachurik$^\sharp{}^\ddag$}
\maketitle

 \centerline{${}^\sharp$\,Bogolyubov Institute for Theoretical Physics of NAS of Ukraine}
 \centerline{14b, Metrologichna Str., Kyiv 03680, Ukraine}

   \vspace{3mm}
\centerline{${}^\ddag$\,Khmelnytskyi National University}
 \centerline{11, Instytutska Str., 29016 Khmelnytskyi, Ukraine}

\vspace{3mm}
\begin{abstract}
The recently introduced two- and three-parameter ($p,q$)- and
($p,q,\mu$)-deformed extensions of the Heisenberg algebra were
explored under the condition of their connectedness with the
respective nonstandard (other than known ones) deformed quantum
oscillator algebras.  In this paper we show that such connection
dictates certain new $\eta(N)$-pseudo-Hermitian conjugation rule
between the creation and annihilation operators, with $\eta(N)$
depending on the particle number operator $N$. In turn, that leads
to the related $\eta(N)$-pseudo-Hermiticity of the position/momentum
operators, though the involved Hamiltonian is Hermitian.
 Different possible cases are studied, and some interesting
 features implied by the use of such $\eta(N)$-based conjugation
 rule are emphasized.
 \end{abstract}

\vspace{3mm}

\vspace{2mm}

\section{Introduction}
 Non-Hermitian modifications of quantum mechanics
 \cite{Geyer,Cannata,Bender1,Znojil,Mostafa1,Ahmed,Mostafa2,Bender2,Bender3,Mostafa,Li,Bagchi}
 which lead nevertheless to real spectra of operators, attract great interest.
 A significant part of such investigations in recent years has been crystalized into
 an important branch encompassing the works on pseudo-Hermitian \cite{Mostafa1}
 representation in quantum mechanics, see the comprehensive review \cite{Mostafa}
 which gives a plenty of references and discusses main ideas and results.
 This approach have its impact on a variety of applications, ranging
from nuclear physics and quantum field theory to nonlinear optics
and biophysics \cite{Mostafa}.

On the other hand, significant attention is devoted to generalized
       versions \cite{Saaved,Brod,Jan,Kempf,Ch-K,Plyush,Leiva,Bagchi,GK-1}
 of Heisenberg algebra (HA), obtained through appropriate
 extension of its basic relation $ [X, P] = {\rm i}\hbar$.
  That implies respective modifications of the uncertainty relation
                        (see e.g. \cite{Jan,Kempf,Garay,Hossen,Dorsch,Tkachuk}).
 In our recent paper \cite{GK-1}, the so-called two-sided or
 three-parameter ($p,q,\mu$)-deformation of HA has been introduced and studied.
    Its particular $p=1, \mu=0$ case yields simple modified HA with $q$-commutator
   involved  which was studied in \cite{Ch-K}, where the explicit relation with
   certain non-standard $q$-deformed oscillator algebra was found
   (for some recent applications of deformed oscillators or deformed bosons
   see e.g. \cite{Chang,Liu,AdGa,SIGMA,GKM,GM1,GM2,GM3}).
   In the more general "left-handed" + "right-handed" generalization \cite{GK-1},
   the analogous mapping onto deformed oscillator has been derived
   in the case of $p,q$-deformation of HA (involving $p,q$-commutator)
   as well as for the case of three-parameter $p,q,\mu$-deformed HA,
   where the related non-standard deformed oscillator algebra (DOA)
   was obtained in a somewhat restricted situation.
    In conjunction with the mentioned relation, an important property was
    deduced that the deformation parameter $\mu$ and also the parameters
    $p$ and $q$ explicitly depend on the particle number operator $N$.

 In all three mentioned cases, i.e. $q$- , ($p,q$)- and ($p,q,\mu$)-deformations
 of HA, the formulas relating the position and momentum operators $X, P$ and
 creation/destruction operators $a^+, a^-$ are not those of usual harmonic
 oscillator, see e.g. \cite{Davydov}, but involve $N$-dependent coefficients.
 This fact, rooted in the imposed condition of realizability
 of particular deformed HA through respective DOA, is of basic importance as it
 dictates principal distinction from the usual (Hermitian) conjugation rules
 of the operators involved. However, such rules were not considered in \cite{GK-1}.

 Therefore, the goal of the present paper is to examine those aspects
 of 2- and 3-parameter deformed HA (or DHA) which concern modified rules of
 (self)conjugation of the operators involved.  In our study, most important
 aspect is the encountered unusual $\eta(N)$-pseudo-Hermitian
 conjugation and the related $\eta(N)$-pseudo-Hermiticity of $X$ and/or $P$.
 About the plan of our paper:  Sections~2-5 give a sketch of those
 deformed versions of HA which serve as playground.
  The reasons of why, instead of usual Hermitian conjugation and usual Hermiticity,
  there inevitably emerges the concept of $\eta(N)$-pseudo-Hermitian conjugation,
  along with $\eta(N)$-pseudo-Hermiticity, are described in Sec.~6.
  Therein, we also study the cases with partial $\eta(N)$-pseudo-Hermiticity
  (when one of the operators, $X$ or $P$, remains Hermitian).
    In Sec.~7 we show that both $X$ and $P$ should obey $\eta(N)$-pseudo-Hermiticity
  even in the case when $a^+$ and $a^-$ are usual Hermitian conjugates of each other.
  General situation when $\eta(N)$-pseudo-Hermitian conjugation concerns all the four
  operators is treated in Sec.~8. Next 9th Section deals with the properties of how
  $X$, $P$ commute with the particle number operator $N$, while the Hamiltonian
  (in terms of $X$, $P$) and its Hermiticity are the subject of Sec.~10.
  The paper ends with concluding remarks.

  \vspace{-2mm}

\section{Extended Heisenberg algebra with Hamiltonian or $P^2$ in R.H.S.}

  The Heisenberg algebra (HA), based on the well-known relation of commutation
 \begin{equation}                 \label{HA}
 [X, P] = {\rm i}\hbar \ ,
\end{equation}
  \vspace{-1mm}
 {\color{black}
during last decades serves as starting point for diverse
modifications or genera\-li\-}zations.
 Rather general and one of most natural extensions of (\ref{HA})
involves in its r.h.s.
  a function $f({\cal H})$ of the Hamiltonian ${\cal H}$, that is
\begin{equation}                    \label{f-HA}
[X, P] = {\rm i} \hbar f({\cal H})\ .
\end{equation}
  Some versions of this modification of HA  were studied e.g. in \cite{Saaved,Brod,Leiva},
  exploiting either $f(P^2,X^2)$, or $f(P^2)$,
  or the particular form $\exp{(\kappa P^2)}$.
 With constant term only (or zeroth order in ${\cal H}$),
the relation (\ref{f-HA}) reduces to the customary HA (\ref{HA}).

 An important special case of Eq. (\ref{f-HA}), namely the algebra
 based on the relation
 \begin{equation}                              \label{mu-HA}
[X, P] = {\rm i} \hbar (1 + \mu {\cal H})\ ,
  \hspace{8mm}  \mu\in\mathbb{R} ,
\end{equation}
 was explored in \cite{Saaved}
  with the impact on quantum mechanics at the extreme
 conditions of high energy physics and quark physics.
 This line of research was developed in a number of papers
                           such as \cite{Brod,Jan,Kempf,Leiva,Bagchi,GK-1}
 and others.


 \section{ A $p,q$-deformed Heisenberg algebra:  the connection with DOA }

 Another approach to deform HA affects the l.h.s.
 of defining relation, and yields the two-parameter or $p,q$-deformation of the form
\begin{equation}                          \label{pqHA}
    p X P-q PX = {\rm i}\hbar \ ,
\end{equation}
introduced and studied in Ref.~20. Note that its special case $p=1$
was earlier analyzed by Chung and Klimyk in Ref. \cite{Ch-K}.

 In Ref. \cite{GK-1}, main goal was to connect
 the deformed HA (\ref{pqHA}), and its two-sided 3-parameter
extension, with an appropriate DOA. As explained in many papers, see
e.g. the overviews \cite{Melj,Bona}, each version of DOA is
generated by three generating elements $a^+$, $a^-$ and $N$ (with
physical meaning respectively the creation, the annihilation, and
the excitation number operators, within the Fock type
representation).
 The three generators of DOA obey the following defining relations:
\begin{equation}                                 \label{N_a^pm}
 [N,a^{\pm}]=\pm a^{\pm} ,  \hspace{10mm}
      a^+a^- =\phi(N) , \hspace{8mm} a^-a^+ =\phi(N+1) ,
\end{equation}
and the generalization
\begin{equation}                               \label{a_phi}
 a^-a^+ - a^+a^- =\phi(N+1) -\phi(N)
\end{equation}
of usual commutation relation $[a^-,a^+]=1$ (this is recovered at
$\phi(N)=N$).
  Here $\phi(N)$ is the {\it deformation structure function}
  (DSF): \ indeed, its  particular form completely defines the corresponding DOA.
  Moreover, in the Fock type representation (and basis) the DSF $\phi(N)$
explicitly determines matrix elements of the operators $a^+ $ and
$a^-$ according to the formulas
\[ N  |n\rangle = n |n\rangle,  \hspace{8mm}
     a^+|n\rangle =\sqrt{\phi(n+1)} |n+1 \rangle,  \hspace{8mm}
     a^-|n\rangle =\sqrt{\phi(n)}|n-1 \rangle .
\]
These explain the names "creation, annihilation, and number"
operators.

 The procedure of connecting the DHA given by (\ref{pqHA}) with an appropriate DOA
 was described in detail in Ref. \cite{GK-1}, so we do not need to reproduce its
 full content here.

 However, some of the results necessary for our present goals will be
 recalled below. Let us first note that there exists yet another form
   of generalized commutation relation for $a^-$ and $a^+$, namely
 \begin{equation}                                   \label{G-H}
G(N) a^-a^+ -  H(N) a^+a^- = 1 .
 \end{equation}
 As known\cite{Melj}, given the latter form one can go over
 to the form (\ref{a_phi}) above involving the DSF $\phi(N)$.

 To find $G(N)$ and $H(N)$ we assume the relation between $X$, $P$
 appearing in (\ref{pqHA}), and the triple $a^-$, $a^+$ and $N$, generating DOA,
 in rather general form
 \[
 X = f(N)\, a^-  +  g(N)\, a^+  ,  \hspace{14mm}     P = {\rm
   i}\bigl(\,k(N)\, a^+ - h(N)\,a^-\bigr)  .
 \]
Using the latter jointly with relations (\ref{pqHA}) and
(\ref{G-H}), one is able to find explicitly (see Ref. \cite{GK-1}
for more details) first the functions $f(N)$, $ g(N)$, $h(N)$,
$k(N)$ and then, by means of these, also the functions $G(N)$ and
$H(N)$:
\[
 f(N) = k(N) = \frac{1}{\sqrt{2}}\, Q^N ,   \hspace{8mm}
 h(N)= g(N) = \frac{1}{\sqrt{2}}\, Q^{2N} ,  \hspace{8mm}   Q\equiv q/p  ;
 \]
\[
 H(N) = \frac12 q\, Q^{2N} \bigl( 1 +  Q^{2N+2} \bigr) ,   \hspace{14mm}
 G(N)=  \frac12 p\, Q^{2N} \bigl( 1 +  Q^{2N-2} \bigr)  .
 \]
 Thus, we infer the important fact that $X$ and $P$ in terms of the creation,
 annihilation, and particle number operators are expressed as
 \begin{equation}
 \vspace{-1mm}                                    \label{X_vs_a}
X = \frac{1}{\sqrt 2} \, \Bigl[Q^{2N}\!a^+
 +  Q^N\!a^- \Bigr]  ,  \hspace{9mm}
 P = \frac{\rm i}{\sqrt 2} \,
 \Bigl[
Q^N\!a^+ -
  Q^{2N}\!a^-\Bigr]  ,   
\end{equation}
   which will be  used in Sections 6-8. \
 The inverse relations readily follow, so that
\begin{equation}
 \vspace{-1mm}                                    \label{a_vs_X}
a^-\!=\! d_{N,Q}\bigl(Q^{-N}X\!+ {\rm i} P\bigr)   , \hspace{5mm}
 a^+\!=\! d_{N,Q}\bigl(X\!- {\rm i} Q^{-N} P\bigr), \hspace{5mm}
d_{N,Q}\equiv\sqrt2 (1+Q^{2N})^{-1}.
\end{equation}
 Obviously, the restriction $Q=1$ implies $d_{N,1}=\frac{1}{\sqrt2}$ and
 brings us back to the well-known linear relations between  $a^+, a^-$ and $X, P$
 (see e.g. \cite{Davydov}).

\section{Skew-Hermiticity of the basic relation (\ref{pqHA})}
 \medskip

Here we examine consistency of the basic relation (\ref{pqHA}) from
the viewpoint of conjugation: since r.h.s. of (\ref{pqHA}) is
skew-Hermitian, the same property should be valid for the l.h.s.
 To this end, consider the cases of real and complex $p,q$ separately.

\vspace{1mm}
  (A) \underline{Let $p,q \in \mathbb{R}$}.
  \vspace{2mm}

 \noindent Assume that $X^\dag = X$ and $P^\dag = P$.
 Then, skew-hermiticity of the l.h.s. of (\ref{pqHA}) does holds
 only if $p=q$.  This case however is not interesting for us as it
 reduces to the non-deformed one for the operators $\tilde{X}$ and $\tilde{P}$
 such that $\tilde{X}\equiv\sqrt{q}\, X$ and $\tilde{P}\equiv\sqrt{q}\, P$.

 Now let $P^\dag = P$, but $X^\dag =\kappa X \neq X$ with constant $\kappa$.
 By demanding skew-hermiticity of the l.h.s. of (\ref{pqHA}) we deduce:
\[
(\kappa p - q) P X + (p -\kappa q) X P = 0 .
\]
For $P X \neq 0$, it must be that either
 (i) $q= \kappa p$ and $p= \kappa q$  which implies $\kappa^2 =1$
 i.e. $\kappa = \pm 1$ (the first option is trivial and the second one is unphysical),
 or (ii) there should be  $ X P = \omega  P X$
 where $\omega = \frac{\kappa p -q}{\kappa q -p}$. Using (\ref{pqHA}) we
 infer that $P X =  {\rm i}\hbar\, I / (p\omega -q) $ which means $P$ is
 proportional to inverse of $X$, which is also rather exotic.

Same conclusion is drawn if $X^\dag = X$ and $P^\dag = \kappa P$, or
if $X^\dag =\kappa X$ and $P^\dag =\kappa' P$.

More general case involves  $P^\dag = P$ and
 $X^\dag =\tilde{\eta} X \tilde{\eta}^{-1}$ (i.e. the operator $X$ is
 pseudo-Hermitian), and the case with both $P^\dag = {\eta}' P ({\eta}')^{-1}$
 and $X^\dag =\tilde{\eta} X \tilde{\eta}^{-1}$ (the two operators are
 pseudo-Hermitian).
  This will be considered below, see Sec. 7.

\vspace{2mm}
  (B) \underline{Let $p,q \in \mathbb{C}$}.
\vspace{2mm}

\noindent With same assumption that $X^\dag = X$ and $P^\dag = P$,
we infer:
\[
(\bar{p} -q) P X + (p - \bar{q}) X P =0.
\]
For $X P \neq 0$, we have that either

(i) $p=\bar{q}=r{\rm e}^{-{\rm i}\theta }$ and thus ${\rm e}^{-{\rm
i}\theta}X P - {\rm
e}^{{\rm i}\theta} P X =\frac{{\rm i}\hbar}{r} $,  
or

(ii) $p\neq\bar{q}$ and then $P X = \frac{\bar{q}-p}{\bar{p}-q} X
P$, or equivalently $[P,X]_{\tilde{Q}}=0$ where $[A,B]_s\equiv
AB-sBA$
 and $\tilde{Q}\equiv\frac{\bar{q}-p}{\bar{p}-q} =
  -\frac{p-\bar{q}}{(p-\bar{q})^*}= -{\rm e}^{2{\rm i}\arg(p-\bar{q})} $.

 The found restrictions concern parameters $p, q$.
  In general (and more realistic in presence of deformation) case we
  will deal with pseudo-Hermitian $X$ and/or $P$.

 \section{Two-sided (or three-parameter) deformed Heisenberg algebra}
 \medskip

The two-sided, 3-parameter deformed extension of HA recently
introduced in Ref. \cite{GK-1}
 combines different modifications of the HA (\ref{HA}) that yields
\begin{equation}                                 \label{qp-H}
  p X P -q P X = {\rm i} \hbar (1 + \mu {\cal H})\ .
\end{equation}
 Again it is linked with certain deformed boson algebra such that
 the two relations
\begin{equation}                                           \label{31}
   \tilde{H}(N)a^- a^+ - \tilde{G}(N)a^+ a^- = 1 ,
  \hspace{10mm}
a^- a^+ - a^+ a^- = \tilde{\phi}(N+1) - \tilde{\phi}(N)
\end{equation}
  are valid, where $\tilde{\phi}(N)$ is the respective
  structure function of deformation \cite{Melj,Bona}.
 This DSF was derived, in terms of the found  $\tilde{H}(n)$
 and $\tilde{G}(n)$, for two important cases \cite{GK-1}:

 (i)  If $\mu=0$ is set in (\ref{qp-H}) (turning it into (\ref{pqHA})), the proper DSF
      in (\ref{31}) results as
\[  
    \hspace{14mm}
\tilde{\phi}(n) \!=\! \frac{2 p^{-1} Q^{-n}}{(1+Q^{2n-2})
(1+Q^{2n})}
 \left(1\!+\!\frac{Q^{n}\!-\!
  Q^{-n+1}}{Q-1}\right) =
  \] 
\begin{equation}                                           \label{Phi'(n)}
 \ \ \ =\! \frac{2 q^{-n} p^{5n-3}}
                {(q^{2n-2}+p^{2n-2})(q^{2n}+p^{2n})}
 \left(1\!+\!\frac{[2n\!-\!1]_{q,p}}{(qp)^{n-1}}\right)
\end{equation}
(here $[m]_{q,p}\equiv \frac{q^m-p^m}{q-p}$ denotes the $q,p$-number
corresponding to a number $m$), and

(ii) for $\tilde{H}(N)=\tilde{G}(N)$ at $p\neq q$ we obtain (denote
$Q=p/q$):
\[\breve{\phi}(n)=\frac{4Q^2}{p(1+Q^2)(1+Q^3)}
- \frac{4}{p(1+Q)} \biggl(\frac{1-Q^{2-2n}}{1-Q^2}\biggr.
\biggl. + \sum^{n-1}_{j=1}\frac{1+Q^5}{Q^2(1+Q)+Q^{2j}(1+Q^5)}
\biggr).
\]
 Recall that each DSF (e.g. $\tilde{\phi}(N)$)
 relates $a^+ a^-$, $ a^- a^+$ and $N$
 according to formulas
\[
a^+ a^- = \tilde{\phi}(N), \ \ \ \ \ a^- a^+ = \tilde{\phi}(N+1) ,
\]
and determines the corresponding action
 formulas for $a^+$,\ $a^-$ in the normalized basis of
 deformed analog \cite{Bona}  of Fock space so that
\[
 a^\pm \,|n\rangle = \sqrt{\tilde{\phi}\Bigl(n+\frac{1\pm1}{2}\Bigr)} \,|n\pm1\rangle ,
    \hspace{9mm}
 |n\rangle =\bigl(\tilde{\phi}(n)!\bigr)^{-\frac12} (a^+)^n |0\rangle ,
   \hspace{7mm}   a^-|0\rangle = 0 ,
\]
where $\tilde{\phi}(n)! = \tilde{\phi}(n)\,\tilde{\phi}(n\!-\!1)
\ldots \tilde{\phi}(2)\,\tilde{\phi}(1)$.

  Formula (\ref{Phi'(n)}) gives the DSF of \underline{nonstandard}
two-parameter deformed quantum oscillator.
 Nonstandard means it is nonsymmetric under $q\leftrightarrow p$ because
 of the factor $q^{-n}p^{5n-3}$ in the numerator.
 Thus it obviously differs from the well-known $q,p$-oscillator \cite{Chakra}
 whose structure function $\varphi_{q,p}(n)=[n]_{q,p}$ is
  ($q\leftrightarrow p$)-symmetric.

Let us note that formulas (\ref{X_vs_a})-(\ref{a_vs_X}) and the
conclusions in the preceding Section about skew-Hermiticity extend
to the two-sided deformation of HA, see Eq.~(\ref{qp-H}), under the
condition that $\mu$ is real and the Hamiltonian $\cal{H}$ is
Hermitian (the Hermiticity of $\cal{H}$ is discussed in Sec.~9
below).

 \section{An $\eta(N)$-pseudo-Hermitian conjugation of the operators $a^+$,  $a^-$}

In this Section, two distinct cases will be considered.

\underline{Case A}. Assume, at Hermitian $N$, the Hermiticity for
the momentum operator
\begin{equation}              \label{P_hermit}
 P^\dagger = P  \, ,
\end{equation}
and then infer conjugation rules for $a^{\pm}$.
 From Eq.~(\ref{P_hermit}), using Eq.~(\ref{X_vs_a}) we have
 \[ P^\dagger = \frac{\rm -i}{\sqrt 2} \,
 \Bigl[
(a^+)^\dagger Q^N\! -
  (a^-)^\dagger Q^{2N}\!\Bigr]=
  \frac{\rm i}{\sqrt 2} \,
 \Bigl[
Q^N\!a^+ -
  Q^{2N}\!a^-\Bigr]=\frac{\rm i}{\sqrt 2} \,
 \Bigl[a^+ Q^{N+1}\! -
  a^-Q^{2N-2}\! \Bigr]
\]
that yields:
 \vspace{-3mm}
\[ 
(a^+)^\dagger Q^N  = a^- Q^{2N-2} , \ \ \ \
  (a^-)^\dagger Q^{2N}= a^+ Q^{N+1} .
\] 
 From this, using the relation (\ref{F-a_pm}) below,
 we infer \underline{the new conjugation rules}:
\begin{equation}          \label{a_pseudo}
(a^+)^\dagger  =  \eta(N) a^- , \hspace{8mm}
  (a^-)^\dagger = a^+ \eta^{-1}(N),  \hspace{15mm}  \eta(N)\equiv
  Q^{N-1} .
\end{equation}
We call this new kind of conjugation $\eta(N)$-{\it pseudo-Hermitian
conjugation}: it generalizes to $\eta=\eta(N)$ the known
      $\eta$-pseudo-Hermitian conjugation \cite{Mostafa,Li,Bagchi}
  (note that $\eta$ in those papers depended on the momentum $P$).
 Thus, $a^+$ and $a^-$ are mutual $\eta(N)$-{\it pseudo-Hermitian}
 conjugates of each other.

 It is clear that instead of (\ref{a_pseudo}) we can also adopt
 the equivalent definition of
  $\eta(N)$-{\it pseudo-Hermitian conjugation}, namely
\begin{equation}          \label{a_pseudo'}
(a^+)^\dagger  =  a^- \eta(N) , \hspace{8mm}
  (a^-)^\dagger = \eta^{-1}(N) a^+ ,  \hspace{15mm}  \eta(N)\equiv Q^{N-2}.
\end{equation}
 Obviously, when $Q\to 1$ (i.e. at $p=q$), the both versions of
 $\eta(N)$-pseudo-Hermitian conjugation, (\ref{a_pseudo}) and (\ref{a_pseudo'}), go over
 into the usual Hermitian mutual conjugation of $a^+$ and $a^-$.
 We stress that this concerns the $p$, $q$ differing from unity, as well as
 when the both are equal to $1$.

  \underline{Remark 1}. With account of (\ref{a_pseudo}), we have usual Hermiticity
  for the bilinears,
\begin{equation}          \label{a+a-}
(a^+ a^-)^\dagger  = (a^-)^\dagger (a^+)^\dagger=
a^+\eta^{-1}(N)\eta(N)a^- = a^+ a^- ,
\end{equation}
\begin{equation}          \label{a-a+}
(a^- a^+)^\dagger = (a^+)^\dagger (a^-)^\dagger= \eta(N) a^-
  a^+\eta^{-1}(N) = a^- a^+ ,
\end{equation}
 where at the last step in (\ref{a-a+}) the permutation rule
\begin{equation}          \label{F-a_pm}
  {\cal F}(N) a^{\pm}=a^\pm {\cal F}(N\pm 1)\, ,
\end{equation}
stemming from the first relation    
 in (\ref{N_a^pm}) and valid for general function ${\cal F}(N)$, has
been utilized.
 The same is true if one takes (\ref{a_pseudo'}).

 \underline{Remark 2}.  Instead of (\ref{a_pseudo}) (resp. (\ref{a_pseudo'}))
 we of course could take the standard, as for operators,
 shape of mutual conjugation, i.e.  
$(a^+)^\dagger = \zeta(N) a^- \zeta^{-1}(N)$ and $(a^-)^\dagger =
\zeta(N) a^+ \zeta^{-1}(N)$. However in view of (\ref{F-a_pm}),
after redefinition $\zeta(N-1)\zeta^{-1}(N)\to \eta(N)$ (resp.
$\zeta(N)\zeta^{-1}(N+1)\to \eta(N)$), that would reduce to
(\ref{a_pseudo}) (resp. (\ref{a_pseudo'})).

 \vspace{3mm}
  {\it Pseudo-Hermiticity of the position operator $X$}
 \vspace{3mm}

Recall that $\eta(N)$-{\it pseudo-Hermitian conjugation}
(\ref{a_pseudo}) of $a^+$ and $a^-$ has been inferred in view of the
requirement (\ref{P_hermit}).
  On the other hand, the property (\ref{a_pseudo}) causes
  non-Hermiticity of the operator $X$.
 That is, we have to modify conjugation rule for the operator $X$.
 So let us find the modified rule of self-conjugation for the position
operator in the assumed form
 $ X^\dagger=\tilde{\eta}^{-1}(N) \, X \,\tilde{\eta}(N)$.
  From (\ref{a_pseudo}) we have
\[
X^\dagger=\frac{1}{\sqrt 2}\Bigl( (a^+)^\dagger Q^{2N}+
(a^-)^\dagger Q^{N}\Bigr)= \frac{Q}{\sqrt 2} \Bigl( a^+ +
Q^{3N}a^-\Bigr) ,
\]
and it can be easily verified that
\begin{equation}               \label{16}
 X^\dagger=\tilde{\eta}^{-1}(N) \, X \,\tilde{\eta}(N) \hspace{7mm}
 {\rm with} \hspace{7mm}
 \tilde{\eta}(N)=Q^{N^2}.
\end{equation}
The same does follow if we take the rule of $\eta(N)$-conjugation in
the form (\ref{a_pseudo'}).

Thus, for the conjugation properties of the momentum and position
operators here we have usual {\it Hermiticity} of $P$, but the
$\tilde{\eta}(N)$-{\it pseudo-Hermiticity of} $X$, i.e.
\begin{equation}                            \label{P_self}
P^\dagger =P    \hspace{12mm}  {\rm and}   \hspace{12mm}
X^\dagger=Q^{-N^2} X \,Q^{N^2} \ .
\end{equation}
That is certainly linked with the rule of $\eta(N)$-{\it
pseudo-Hermitian} mutual conjugation for $a^+$ and $a^-$ given by
(\ref{a_pseudo}) or (\ref{a_pseudo'}).

Similar analysis can be carried out if one exchanges the roles of
$X$ and $P$.

\underline{Case B}.  This time let us require that
 \begin{equation}
 X^\dagger = X.
  \end{equation}
 Then we are led to the conjugation rule
 \begin{equation}                                    \label{a_pseudo''}
 (a^+)^\dagger = \hat{\eta}(N) a^- , \hspace{7mm}
 (a^-)^\dagger = a^+ \hat{\eta}^{-1}(N) , \hspace{10mm} \hat{\eta}(N)=
 Q^{-N-2} .
 \end{equation}
 As a consequence we arrive at $\tilde{\tilde{\eta}}(N)$-pseudo-Hermiticity
 of $P$, i.e.
  \begin{equation}
  P^\dagger = \tilde{\tilde{\eta}}^{-1}(N) \, P\,
  \tilde{\tilde{\eta}}(N)    \hspace{7mm}  {\rm with} \hspace{7mm}
  \tilde{\tilde{\eta}}(N) = Q^{-N^2}.
  \end{equation}
Thus, for the (self)conjugation rules for momentum/position
operators in this case we have usual {\it Hermiticity} of $X$
jointly with $\tilde{\tilde{\eta}}(N)$-{\it pseudo-Hermiticity of}
$P$:
 \begin{equation}                                  \label{X_self}
X^\dagger =X    \hspace{12mm}  {\rm and}   \hspace{12mm}
P^\dagger=Q^{N^2} P \,Q^{-N^2} ,
\end{equation}
the both linked with $\hat{\eta}(N)$-{\it pseudo-Hermitian
conjugation} of $a^+$ and $a^-$ in (\ref{a_pseudo''}).

 It is interesting to compare the coordinated couple of conjugation rules
 (\ref{a_pseudo}) and (\ref{P_self}),  with the respective coordinated couple
 of conjugation rules (\ref{a_pseudo''}) and (\ref{X_self}).

   \underline{Remark 3}.  It should be stressed that the ${\eta}(N)$-dependence
   in the conjugation  rules for $a^+$ and $a^-$, and the related
   ${\eta}(N)$-self-conjugation of  $X$ and/or $P$, are rooted in the basic
   connection established in Ref.~20 :\
   \underline{DHA $\Leftrightarrow$ DOA}
   (i.e. deformed Heisenberg algebra $\Leftrightarrow$ deformed oscillator algebra).

 \section{The case when $a^+$ and $a^-$ are usual conjugates of
 each other}

 Now let us require for $a^+$ and $a^-$ the usual conjugation property: \
$(a^\pm)^\dagger =  a^\mp$. Then it is easy to see that both
$X^\dagger\neq X$ and $P^\dagger\neq P$.
 Therefore we consider these operators as $\eta(N)$-pseudo-Hermitian ones,
 by imposing
\begin{equation}
X^\dagger = \eta^{-1}_{\scriptstyle{X}}(N) \, X \,
\eta^{}_{\scriptstyle{X}}(N) \, ,  \hspace{10mm}
  P^\dagger = \eta^{-1}_{\scriptstyle{P}}(N)   \, P \, \eta^{}_{\scriptstyle{P}}(N)
\end{equation}
where $\eta^{}_{\scriptstyle{X}}(N)$ and
$\eta^{}_{\scriptstyle{P}}(N)$ are some functions of $N$ possessing
their corresponding inverses.
 To find $\eta^{}_{\scriptstyle{X}}(N)$ and $\eta^{}_{\scriptstyle{P}}(N)$
 explicitly, we use the formulas (\ref{X_vs_a}) for $X$ and $P$.
 Then, by a simple algebra we deduce the following recurrence relations:
\[
\eta^{}_{\scriptstyle{X}}(N+1)=\eta^{}_{\scriptstyle{X}}(N)\,
Q^{N+2} \, , \hspace{10mm}
\eta^{}_{\scriptstyle{P}}(N+1)=\eta^{}_{\scriptstyle{P}}(N)\,
Q^{-N+1}\, .
\]
Solving them we find respectively
\[
\eta^{}_{\scriptstyle{X}}(N) = \
 Q^{\frac12 N(N+3)}\, \eta^{}_{\scriptstyle{X}}(0)\, , \hspace{10mm}
 \eta^{}_{\scriptstyle{P}}(N) =
\, Q^{\frac12 N(-N+3)}\, \eta^{}_{\scriptstyle{P}}(0) .
\]
Obviously, the convenient choice is to set
$\eta^{}_{\scriptstyle{X}}(0)= \eta^{}_{\scriptstyle{P}}(0)=1$.

\section{On the $\eta(N)$-pseudo-Hermitian conjugation of $a^\pm$, $X$ and $P$}

To consider most general situation when the rules of
pseudo-Hermitian conjugation concern both the pair $a^+$,\ $a^-$ and
the operators $X$,\  $P$, we impose the relations
 \begin{equation}   \label{eta_a}
 ( a^+)^\dagger = \eta_a(N)\, a^- \, ,
 \hspace{10mm}  ( a^-)^\dagger = a^+ \, \eta^{-1}_a(N) \, ,
 \end{equation}
 \begin{equation}    \label{eta_x,eta_p}
  X^\dagger = \eta^{-1}_X(N) \, X \, \eta^{}_X(N) \, , \hspace{11mm}
  P^\dagger = \eta^{-1}_P(N) \, P \, \eta^{}_P(N) \, ,
 \end{equation}
where all the three $\eta$'s are different.

We wish to find relations governing the $\eta$'s. For this, we take
conjugate $X^\dagger$ of $X$ in (\ref{X_vs_a}), then use
(\ref{eta_a}) and compare with $X^\dagger$ in (\ref{eta_x,eta_p}).
 That results in the equations
\[
Q^{2N+2}\, \eta_a(N) = Q^{N}
 \frac{\eta^{}_X(N\!+\!1)}{\eta^{}_X(N)}  \, ,
 \hspace{14mm}    
\frac{Q^{N-1}}{\eta_a(N\!-\!1)} = \frac{Q^{2N}
 \eta^{}_X(N\!-\!1)}{\eta^{}_X(N)} \, ,
\]
or equivalently in the equations
\begin{equation}           \label{eta_a,x}
\eta_a(N) = Q^{-N-2}
 \frac{\eta^{}_X(N\!+\!1)}{\eta^{}_X(N)}  \, ,
 \hspace{14mm}
\frac{1}{\eta_a(N\!-\!1)} = \frac{Q^{N+1}
 \eta^{}_X(N\!-\!1)}{\eta^{}_X(N)} \, .
\end{equation}
 The latter two are not independent, being inverse
of each other (shift $N\to N\!+\!1$).

Likewise, taking conjugate of $P$ in (\ref{X_vs_a}), then using
(\ref{eta_a}) and comparing with $P^\dagger$ in (\ref{eta_x,eta_p}),
 we obtain the equations
\[
Q^{N+1}\, \eta_a(N) = Q^{2N}
 \frac{\eta^{}_P(N\!+\!1)}{\eta^{}_P(N)}  \, ,
 \hspace{14mm}    
\frac{Q^{2N-2}}{\eta_a(N\!-\!1)} = \frac{Q^{N}
 \eta^{}_P(N\!-\!1)}{\eta^{}_P(N)} \,  \, ,
\]
or equivalently the equations
 \begin{equation}     \label{eta_a,p}
\eta_a(N) = Q^{N-1}
 \frac{\eta^{}_P(N\!+\!1)}{\eta^{}_P(N)}  \, ,
 \hspace{14mm}   
\frac{1}{\eta_a(N\!-\!1)} = \frac{Q^{-N+2}
 \eta^{}_P(N\!-\!1)}{\eta^{}_P(N)} \,  .
\end{equation}
 Again the latter two are not independent, but inverse
of each other.

At last, from (\ref{eta_a,x}) and (\ref{eta_a,p}) by excluding
$\eta_a$ we infer the relation connecting $\eta^{}_X(N)$ with
$\eta^{}_P(N)$, namely
 \begin{equation}         \label{eta_x,p}
\frac{\eta^{}_X(N\!+\!1)}{\eta^{}_X(N)} = Q^{2N+1}
 \frac{\eta^{}_P(N\!+\!1)}{\eta^{}_P(N)} \, .
 \end{equation}
  Thus, for finding $\eta_a(N)$, $\eta^{}_X(N)$ and $\eta^{}_P(N)$
  we have three relations: that is Eq.(\ref{eta_x,p})
  and, say, the first ones in (\ref{eta_a,x}), (\ref{eta_a,p}), so
  that any two of the three are independent.

  Now let us examine different possible situations.

 \vspace{1mm}
  (i) It follows from (\ref{eta_x,p}) that $\eta^{}_X(N)\ne
  {\rm const} \cdot \eta^{}_P(N)$ for any $Q\ne 1$.

 \vspace{1mm}
(ii) If  $\eta^{}_X(N)$ is known (or chosen), then $\eta_a(N)$
follows explicitly, see (\ref{eta_a,x}), and for $\eta^{}_P(N)$ we
have recursion relation which can be easily solved.

 \vspace{1mm}
(iii) Likewise, if $\eta^{}_P(N)$ is known (or chosen), then
$\eta_a(N)$ follows explicitly, see (\ref{eta_a,p}), and for
$\eta^{}_X(N)$ we have recursion relation which can be easily
solved.

 \vspace{1mm}
 (iv) If  $\eta_a(N)$ is fixed (chosen), then we have two similar,
 though not identical, recursion relations for $\eta^{}_X(N)$
 and $\eta^{}_P(N)$ to be solved.

  \vspace{1mm}

  \hspace{10mm}
  \underline{It is worth to consider some particular cases}:

     \vspace{1mm}
 (a)  Put $\eta_a(N)=Q^{-N-2}$ in (\ref{eta_a,x}).
  Then $\eta^{}_X(N)={\rm const}$ and thus $X$ is Hermitian:  $X^\dagger = X$.
  For $\eta^{}_P(N)$, from recurrence relation (\ref{eta_a,p}) we then find
  $\eta^{}_P(N)=Q^{-N^2}$.

 \vspace{1mm}
 (b)  Put $\eta_a(N)=Q^{N-1}$ in (\ref{eta_a,p}). Then $\eta^{}_P(N)={\rm const}$ and
 thus $P$ is Hermitian:  $P^\dagger = P$. For $\eta^{}_X(N)$, from recurrence relation
 (\ref{eta_a,x}) we then find $\eta^{}_X(N)=Q^{N^2}$.

 \vspace{1mm}
 (c) Put $\eta_a(N)\!=\!1$ that implies $(a^\pm)^\dagger\!=\!a^\mp$ (see also Sec.~6).
  Then from the respective recursion relations we find
   $\eta^{}_X(N)= Q^{\frac12N(N+3)} $ \  and $\eta^{}_P(N)= Q^{\frac12N(-N+3)}$.

 \vspace{1mm}
  (d) Let $\eta_a(N)={\rm const}\ne 1$, for instance $\eta_a =
 Q^{\alpha}$ with real ${\alpha}$. Then for $a^\pm$ we have
standard pseudo-Hermitian conjugation of the shape
 $(a^\pm)^\dagger= Q^{\pm\alpha}a^\mp$.
  The remaining $\eta^{}_X(N)$ and $\eta^{}_P(N)$ are found
  from the relevant recurrence relations, and the result is
  $\eta^{}_X(N)= Q^{\frac12 N (N+3\pm 2\alpha)}$
 and $\eta^{}_P(N)= Q^{\frac12 N (-N+3\pm 2\alpha)}$ .

 \section{Commutation of $X$ and $P$ with the number operator $N$}

For what follows we need the relations of permutation of the number
operator $N$ with the position or momentum operators,
\[
[N,X]= \frac{1}{\sqrt2}(Q^{2N}a^+ - Q^{N}a^-)= X-2Q^N a^-  \, ,
\]
\[
[N,P]= \frac{\rm i}{\sqrt2}(Q^{N}a^+ + Q^{2N}a^-) = P+2{\rm i}
Q^{2N} a^-  \, ,
\]
from which we have
\[
q^{\pm N}[N,X] \mp {\rm i} [N,P] =  -{\rm i} P \pm q^{\pm N} X
\]
and, denoting $q^{\pm N} X\equiv X^{(\pm)}_{N,q}$, infer
\begin{equation}                                 \label{N_XP}
[N,X^{(\pm)}_{N,q}\mp{\rm i}P]= \pm (X^{(\pm)}_{N,q}\mp{\rm i} P)
 \hspace{4mm} \Leftrightarrow  \hspace{4mm}
 N(X^{(\pm)}_{N,q}\mp{\rm i}P) = (X^{(\pm)}_{N,q}\mp{\rm i}P)(N\pm1)
 . \hspace{1mm}
\end{equation}
  It is also possible to infer an interesting relations
( containing $a^-$ explicitly), e.g.
\[
\begin{array}{l}
NX = X(N+1) - \sqrt2 q^N a^- , \\
N^2 X = X(N+1)^2 - \sqrt2 q^N a^- (2N) , \\
N^3 X = X(N+1)^3 - \sqrt2 q^N a^- (3N^2+1) , \\
N^4 X= X(N+1)^4 - \sqrt2 q^N a^- (4N^3+4N)
 ,  \end{array}
\]
and so on.
 It is easily seen that these particular cases generalize to
 \[
N^k X\!=\!X(N\!+\!1)^k \!-\! \sqrt2 q^N a^- A_k(N) \, ,
 \]
where $A_k(N)$ obeys the recurrence formula
\[
A_{k+1}(N)= 2N A_k(N) -(N-1)(N+1)A_{k-1}(N)
 \]
solved by
    \vspace{-2mm}   
 \begin{equation}                       \label{A_k}
 A_k(N)\!=\!\sum^{k-1}_{r=0}
(N\!+\!1)^{k-1-r} (N\!-\!1)^{r}=\frac{(N+1)^k-(N-1)^k}{2} .
\end{equation}
Equivalently,
   \vspace{-2mm}
\begin{equation}                                 \label{NkX}
N^k X= X(N+1)^k - \sqrt2 q^N A_k(N+1) a^- \, .
 \end{equation}
 Using the latter, we arrive at the desired relation involving
 general function ${\cal F}(N)$:
\begin{equation}                             \label{FN_X}
 {\cal F}(N) X = X {\cal F}(N+1) - [{\cal F}(N+2) -{\cal F}(N)]
 \tilde{a}^- \, ,   \ \ \   \tilde{a}^-\equiv \frac{1}{\sqrt2} Q^N
 a^- .
\end{equation}
 Likewise, for the pair $N$ and $P$ we first obtain (compare with (\ref{NkX}))
 \begin{equation}
N^k P = P(N+1)^k +  {\rm i}\sqrt2 q^{2N} A_k(N+1) a^- \ ,
  \end{equation}
with the same $A_k(N)$ as in (\ref{A_k}) above. \vspace{1mm}
Again, from the latter formula we find for general function ${\cal
F}(N)$ the relation
\begin{equation}                                        \label{FN_P}
 {\cal F}(N) P = P {\cal F}(N+1) + [{\cal F}(N+2)-{\cal F}(N)]
 \hat{a}^- \, ,   \ \ \   \hat{a}^-\equiv \frac{\rm i}{\sqrt2}Q^{2N}\, a^- \, .
\end{equation}
Note that for particular ${\cal F}(N)=Q^N$, or $Q^{-N}$ the above
formulas take simpler form:
\[
Q^{\pm N} X = X Q^{\pm (N+1)}\pm (1-Q^2)Q^{\pm (N+1)-1}
\tilde{a}^-\, ,
\]
\[
 Q^{\pm N} P = P Q^{\pm (N+1)}\mp
(1-Q^2)Q^{\pm (N+1)-1} \hat{a}^- \, .
\]
 We see that under the replacement $N \to N\pm 1$ the entities $\tilde{a}^-$ and
 $\hat{a^-}$ in these formulas do not change.
  Remark also that  $a^- \tilde{a}^-=Q \tilde{a}^- a^-$, \ \
  $a^- \hat{a}^-=Q^2\hat{a}^- a^-$.
  Using the above results (\ref{FN_X}) and (\ref{FN_P})
  we deduce the following relation of permutation
\begin{equation}                                \label{FN_XP}
 {\cal F}(N)(X^{(\pm)}_{N,q}\mp {\rm i} P) =
  (X^{(\pm)}_{N,q}\mp {\rm i} P)\,{\cal F}(N\pm1)
 \end{equation}
 for an operator function ${\cal F}(N)$\
 (possessing expansion into a formal series).
 This is nothing but generalization of Eq.~(\ref{N_XP}).

 Let us stress again that the obtained relations of
commutation between $X$, $P$ and (a function of) $N$, see
(\ref{FN_X}), (\ref{FN_P}) and (\ref{FN_XP}), are of importance just
for the chosen (in ref. \cite{GK-1} and herein) line of research
 based on the link: deformed Heisenberg algebra $\Leftrightarrow$ deformed
 oscillator algebra. That will be used in our subsequent work.

 \vspace{1mm}
\section{Hamiltonian in terms of the position and momentum operators}

 Consider first the particular case  $\mu=0$ of the algebra
 (\ref{qp-H}). We use the Hamiltonian taken in the
 conventional form \cite{Bona}
\begin{equation}                                \label{H_by_a}
 {\cal H}=\frac12 (a a^+ + a^+ a)=
  \frac12 \Bigl({\tilde\Phi(N+1)}+{\tilde\Phi(N)}\Bigr)\,
\end{equation}
which yields the energy spectrum $E(n)= \frac12
\bigl({\tilde\Phi(n+1)}+{\tilde\Phi(n)}\bigr)$ in the Fock-like
basis.
  With account of eq. (\ref{a_vs_X}) and recalling that
  $d_{\scriptstyle{N}}\equiv d_{\scriptstyle{N,Q}}\equiv\sqrt2
  \,(1+Q^{2N})^{-1}$,
  we find the Hamiltonian in terms of the position and momentum operators, namely
  \[ \hspace{-36mm} {\cal H} = \frac12 d_{\scriptstyle{N}} Q^{-N}
 \left\{ ( d_{\scriptstyle{N+1}}+ Q d_{\scriptstyle{N-1}})
         (X^2+Q^{-1}P^2)+ \right. \ \ \ \      
\]
\begin{equation}                                \label{H_by_XP}
    \left. \  + \ {\rm i}\ (Q^N d_{\scriptstyle{N+1}}-
                         Q^{1-N}d_{\scriptstyle{N-1}})P X +
 {\rm i}(Q^N d_{\scriptstyle{N-1}}-Q^{-1-N}d_{\scriptstyle{N+1}})
  X P\right\} \, ,
 \end{equation}
 which is somewhat reminiscent of the Swanson model \cite{Swanson}.
  With the use of (\ref{pqHA}) this Hamiltonian takes the form
 \[
{\cal H} = \frac12 d_{\scriptstyle{N}}~Q^{-N}
 \left\{ (d_{\scriptstyle{N+1}}+Q d_{\scriptstyle{N-1}})
  \bigl[X^2+Q^{-1}P^2+ {\rm i}Q^{-1}(Q^N - Q^{-N}) XP + \right.
 \]
\begin{equation}
\left.                           \label{H_by_XP2}
 +(1/{q}) (Q^N d_{\scriptstyle{N+1}}-Q^{1-N}d_{\scriptstyle{N-1}})\bigr]\right\}
 \end{equation}
 with $P X$ term now absent.
  Note that at $p\to 1$ the results obtained here
 for the $p,q$-deformed HA reduce to those of the
 one-parameter case (since (\ref{pqHA}) reduces to the $q$-deformed HA
 considered in Ref. \cite{Ch-K}),
 whereas for the case $Q=1$ and $p=q\neq 1$  we come to  the
 structure function $\phi(n)\!=\!\frac{n}{q}$,
 with $X$ and $P$ the same as those mentioned in the last
 line of Sec.3. Obviously, that again leads to the usual harmonic oscillator,
 whose spacing in the (linear) energy spectrum gets $\frac1q$\,-\,scaled.

\vspace{1mm}
     \underline{\it Hermiticity of the Hamiltonian}
 \vspace{1mm}

 As mentioned our Hamiltonian has the form
 $ {\cal H} =\frac12 \bigl( a^- a^+ + a^+ a^- \bigr)$,
 see (\ref{H_by_a}).
 Recall that the creation and annihilation operators are in general
not Hermitian conjugates of each other but instead satisfy the rules
of generalized mutual $\eta(N)$-pseudo-Hermitian conjugation, see
Eq.~(\ref{a_pseudo}) or Eq.~(\ref{eta_a}).
 However, in view of (\ref{a+a-}) and (\ref{a-a+}) this form of Hamiltonian
 guarantees that it is Hermitian.
 The same is true for (\ref{H_by_XP}) and (\ref{H_by_XP2}) as these
 are related with (\ref{H_by_a}) through simple transformation.

The Hamiltonian ${\cal H}$, with account of the equality
\[
\frac{p}{2} Q^{2N+1} \bigl(1+Q^{2N+2}\bigr) a^-a^+ - \frac{p}{2}
Q^{2N} \bigl(1+Q^{2N-2}\bigr) a^+ a^- = 1  ,
\]
 see Eq.(\ref{G-H}) and the formulas above Eq.(\ref{X_vs_a}),
 can be presented as
\begin{equation}                       \label{H_by_a2}
{\cal H} = \frac{1}{p} \ \frac{Q^{-2N-1}}{1+Q^{2N+2}} +\frac12
 \biggl(1+ Q^{-1}\frac{1+Q^{2N-2}}{1+Q^{2N+2}}
\biggr) a^+ a^- \, .
\end{equation}
 This is still Hermitian, in view of Hermiticity of (an
 operator function of) $N$ and the property (\ref{a+a-}) of $a^+a^-$.
At $p=q$, we have $a^-a^+ - a^+a^- = q^{-1}$ and ${\cal H} =
\frac{1}{2q}+ a^+ a^-$. When $q=1$, the usual harmonic oscillator
 with ${\cal H} = {\cal H}_0 = \frac12 + a^+ a^-$ is recovered.

 \vspace{1mm}
 \underline{Remark 4}. The versions of Hamiltonian ${\cal H}$ given
  in (\ref{H_by_XP}), (\ref{H_by_XP2}) and (\ref{H_by_a2}) are
  equivalent to the initial one (\ref{H_by_a}) and thus as well Hermitian.
  On the other hand, the form of Hamiltonian  $H=\frac12 (X^2+P^2)$,
  i.e. the standard one for harmonic oscillator, is not plausible, being
  neither Hermitian nor pseudo-Hermitian in the deformed case of $Q\neq 1$.
   We can however {\it suggest natural and simple modification} of $H$ given in
   terms of $\eta_X$-pseudo-Hermitian $X$ and $\eta_P$-pseudo-Hermitian $P$:
\begin{equation}                       \label{H_XP_eta}
\tilde{\cal H} = \frac12 \Bigl( (\eta_X)^{-\frac12} X^2
(\eta_X)^{\frac12} +  (\eta_P)^{-\frac12} P^2  (\eta_P)^{\frac12}
\Bigr) .
\end{equation}
Then, with the particular $\eta_X$ and $\eta_P$ e.g. corresponding
to $\eta_a=1$, see case (c) at the end of Sec.~8, we obtain
\begin{equation}                       \label{H_XP_QN}
 \tilde{\cal H} =
 \frac12 \Bigl( Q^{-\frac14 N(N+3)} X^2 \, Q^{\frac14 N(N+3)} +
 Q^{\frac14 N(N-3)} P^2 \, Q^{-\frac14 N(N-3)} \Bigr) \ .
\end{equation}
 One can easily check Hermiticity of (\ref{H_XP_eta}) and (\ref{H_XP_QN}).
 Note that if $Q\to 1$, then $\tilde{\cal H} \to H =\frac12(X^2 + P^2)$.

 \vspace{1mm}
 \underline{Remark 5}. Returning to the skew-hermiticity of
 Eq. (\ref{pqHA}), especially its l.h.s., as
discussed in the last part of Sec. 3, we may state the following:
since the Hamiltonian is Hermitian, and $\mu\in \mathbb{R}$, all the
conclusions made at the end of Sec. 3 extend completely to the
(skew-Hermiticity of) {\it three-parameter deformation} of the
 Heisenberg algebra, with its $p,q,\mu$-deformed basic relation Eq. (\ref{qp-H}).

\vspace{-2mm}

\section*{Discussion}

 In the present paper, for the 2- and 3-parameter extensions \cite{GK-1} of
 the Heisenberg algebra, assuming  either usual, or generalized
 (with $\eta_a(N)$ involved) conjugation properties
 of $a^-$ and $a^+$, we studied the special non-Hermiticity of $X$, $P$,
 realized exactly in terms of the notion of  $\eta_X(N)$-pseudo-Hermiticity
 of $X$ or/and $\eta_P(N)$-pseudo-Hermiticity of $P$.
  Generally speaking, our main results concern precise and fully-coordinated
 (mutual or self-) conjugation properties of the four involved operators,
  with the crucial $N$-dependence of the eta-functions $\eta_a$
  and $\eta_X$,\ $\eta_P$.
  Let us stress once again the basical aspect that such $N$-dependence is caused
  by the important link earlier established in Ref. \cite{GK-1}, namely:
   deformed Heisenberg algebra $\Leftrightarrow$ deformed oscillator algebra.
  Also it is worth to note that the "metric" operators $\eta_a(N)$, $\eta_X(N)$ and
  $\eta_P(N)$ are all Hermitian, since they are given as the corresponding functions
  of the Hermitian particle number operator.

  The Hamiltonian ${\cal H}$ in our treatment
 is Hermitian as it is formed from the bilinears
 $a^+ a^-$ and $a^- a^+$.
  Remark that these bilinears are Hermitian, although the
  individual $a^+$ and $a^-$ may be not (mutual) Hermitian conjugates,
   but rather the $\eta_a(N)$-pseudo-Hermitian conjugates of one another.
  In addition we have also introduce the Hamiltonian $\tilde{\cal H}$ as given
  in Eqs. (\ref{H_XP_eta})-(\ref{H_XP_QN}), and this presents yet another
  \underline{Hermitian deformation} of the well-known Hamiltonian
  $H =\frac12(X^2 + P^2)$ of harmonic quantum oscillator.

  In a forthcoming work we intend to examine the spectra
  (eigenvalues, eigenfunctions) of the position and momentum
  operators, along with the Hamiltonian (\ref{H_XP_QN}),
  {\it in the framework of coordinate realization}.
    On the other hand, when exploiting the (deformed) Fock like basis,
    the energy spectrum of the Hamiltonian $H=\frac12\{a^+,a^-\}$ is explicitly
    known: namely it is given, see Eq.~(\ref{H_by_a}), through the respective
    structure function  such as e.g. $\tilde{\Phi}(n)$ in Sec.~4.
    It is also worth to emphasize
  the importance to find and explore particular quantum physical
  systems governed by Hamiltonians such as (\ref{H_by_XP}) and alike, with the
  $\eta(N)$-pseudo-Hermitian position and/or momentum
  operators as those studied above.  Also, it would be interesting to
  compare such results with those obtained in the Swanson model \cite{Swanson}.
  These aspects, along with the study of $\eta(N)$-pseudo-Hermitian Hamiltonians,
  will be among our nearest tasks.

\vspace{-1mm}
\section*{Acknowledgement}

This work was partially supported by the Special Programme of
Division of Physics and Astronomy of NAS of Ukraine.

\end{document}